# CHISL: The Combined High-resolution and Imaging Spectrograph for the LUVOIR Surveyor

Kevin France[a,b], Brian Fleming [a,*], Keri Hoadley[a]

[a]Laboratory for Atmospheric and Space Physics, University of Colorado, Boulder CO 80309, USA
[b]Center for Astrophysics and Space Astronomy, University of Colorado, Boulder CO 80309, USA
[*]NASA Nancy Grace Roman Fellow

## ABSTRACT

NASA is currently carrying out science and technical studies to identify its next astronomy flagship mission, slated to begin development in the 2020s. It has become clear that a Large Ultraviolet/Optical/IR (LUVOIR) Surveyor mission ($d_{primary} \approx 12$ m, $\Delta\lambda \approx 1000$ Å – 2 µm spectroscopic bandpass) can carry out the largest number of NASA's exoplanet and astrophysics science goals over the coming decades. The science grasp of a LUVOIR Surveyor is broad, ranging from the direct detection of potential biomarkers on rocky planets to the flow of matter into and out of galaxies and the history of star-formation across cosmic time. There are technical challenges for several aspects of the LUVOIR Surveyor concept, including component level technology readiness maturation and science instrument concepts for a broadly capable ultraviolet spectrograph. We present the scientific motivation for, and a preliminary design of, a multiplexed ultraviolet spectrograph to support both the exoplanet and astrophysics goals of the LUVOIR Surveyor mission concept, the *Combined High-resolution and Imaging Spectrograph for the LUVOIR Surveyor* (CHISL). CHISL includes a high-resolution ($R \approx 120{,}000$; 1000 - 1700Å) point-source spectroscopy channel and a medium resolution ($R \geq 14{,}000$ from 1000 – 2000 Å in a single observation and $R \sim 24{,}000 – 35{,}000$ in multiple grating settings) imaging spectroscopy channel. CHISL addresses topics ranging from characterizing the composition and structure of planet-forming disks to the feedback of matter between galaxies and the intergalactic medium. We present the CHISL concept, a small sample of representative science cases, and the primary technological hurdles. Technical challenges include high-efficiency ultraviolet coatings and high-quantum efficiency, large-format, photon counting detectors. We are actively engaged in laboratory and flight characterization efforts for all of these enabling technologies as components on sounding rocket payloads under development at the University of Colorado. We describe two payloads that are designed to be pathfinder instruments for the high-resolution (CHESS) and imaging spectroscopy (SISTINE) arms of CHISL. We are carrying out this instrument design, characterization, and flight-testing today to support the new start of a LUVOIR Surveyor mission in the next decade.

## 1. LUVOIR SURVEYOR

The 2010 Astronomy and Astrophysics Decadal Survey (*New Worlds, New Horizons*) recommended an augmentation of the NASA technology development budget for hardware supporting both a mission capable of direct spectroscopy of Earth-like planets and a large UV astrophysics mission to serve as a successor to *HST*[1]. These recommendations built on several exercises for Probe-class (cost ~$1B) missions during the 2000s and numerous white papers making the science case for an Advanced Technology Large Aperture Space Telescope (ATLAST) mission submitted to the 2010 decadal survey for consideration[2]. Additional study led to the inclusion of a "Large Ultraviolet/Optical/InfraRed" (LUVOIR) Surveyor as a Formative Era mission concept in the NASA 2013 Astrophysics Roadmap (*Enduring Quests, Daring Visions: NASA Astrophysics in the Next Three Decades*) for high priority goals for astronomy in the period extending through the 2040s[3].

It was clear from the 2013 roadmap that the LUVOIR Surveyor would yield the largest science return of any of the formative era missions concepts considered, directly addressing the largest number of

kevin.france@colorado.edu

core NASA science priorities in a single mission in the next 30 years, addressing six "primary goals", *twice as many core science goals as any other intermediate-term mission concept* (Section 6.4 in 2013 Roadmap, see also Table 1). The LUVOIR Surveyor directly addresses the primary science goals:

1. Demographics of planetary systems
2. Characterizing other worlds
3. Our nearest neighbors and the search for life
4. The origins of stars and planets
5. The Milky Way and its neighbors
6. The history of galaxies

The LUVOIR Surveyor concept is one of four flagship mission concepts suggested for study by the 2020 Decadal Survey (in addition to the Habitable Planet Explorer (HabEx), an X-ray Surveyor, and a Far-IR Surveyor). By virtue of its ambitious design, LUVOIR carries out the HabEx mission science (topics 1 – 3 above) as well as the suite of general astrophysics priorities (including topics 4 – 6 above). In 2015, NASA's Cosmic Origins Program Analysis Group (COAPG) and Exoplanet Analysis Groups (EXOPAG) both carried out community feedback exercises that supported the inclusion of both LUVOIR and the HabEx concept for study by the 2020 Decadal. In parallel, the Associated Universities for Research in Astronomy (AURA) carried out an independent concept study for a LUVOIR-like mission, the High-Definition Space Telescope (HDST)[4], concluding that this mission could simultaneously carry out the detection and characterization of potentially habitable planets and numerous astrophysical investigations with the large primary aperture and powerful instrument suite envisioned for a LUVOIR-like mission. NASA recently established Science and Technology Definition Teams (STDTs) for these flagship mission concepts.

The exact telescope and instrument capabilities for the LUVOIR Surveyor are not firmly established, but consensus is emerging for a 9.2 – 16 meter primary aperture, likely with a segmented architecture that would deploy on orbit. The COPAG Science Interest Group #2 recommendation for the instrumentation suite for LUVOIR was clear: this mission should be able to carry out the detection and characterization of rocky planets around cool stars and be equipped with wide field imaging and ultraviolet spectroscopy capabilities. This recommendation was taken up in the formal COPAG report to NASA's Science Mission Directorate on flagship missions to be studied by the 2020 Decadal Survey "A flagship mission offering high spatial resolution, high sensitivity, and access to the full range of wavelengths covered by HST (91.2 nm – 2 μm) is essential to advancing key Cosmic Origins science goals in the 2020s and 2030s. Improvement in sensitivity at ultraviolet wavelengths between 91.2 and 110 nm is highly desirable."[5]. In this paper, we describe the concept for a broadly capable UV spectrograph for LUVOIR, the *Combined High-resolution and Imaging Spectrograph for the LUVOIR Surveyor* (CHISL). In Section 2, we describe a small subsample of the exoplanet and astrophysics science cases that could be achieved by LUVOIR with a CHISL-like instrument. Section 3 provides an outline of the CHISL instrument, envisioned as a PI-led instrument contributed to LUVOIR in the tradition of science instruments for *HST* and *JWST*. In Section 4, we describe the technology development ongoing today to make CHISL a technologically viable instrument for a proposal opportunity early in the next decade, including dedicated laboratory and flight testing of CHISL pathfinder instruments on sounding rockets developed at the University of Colorado. Section 5 presents a brief summary of this work.

## 2. ULTRAVIOLET ASTROPHYSICS & EXOPLANET SCIENCE WITH THE LUVOIR SURVEYOR

A persistent theme in astrophysics since the advent of spectroscopy has been the study of gas in the cosmos, its relationship to (and evolution with) star and galaxy formation, and how it is transferred from one site to another. The topic of "gastrophyics" has taken on an even more pervasive role in modern astronomy as we are now characterizing the composition and temperature of gas on scales as large as the cosmic web and as small as the atmospheres of planets around other stars. Understanding the flow of matter and energy from the intergalactic medium (IGM) to the circumgalactic media (CGM), where it can serve as a reservoir for future generations of star and planet formation, is essentially a challenge in characterizing the ionic, atomic, and molecular gas at each phase in this cycle. In the IGM and CGM, the gas is hot and diffuse, best studied in broad atomic hydrogen features (the Lyman series lines of H) and metal lines tracing the temperature range of a few hundreds of thousands of degrees to a few millions of degrees (e.g., O VI and Ne VIII). Once this gas has been accreted into the interstellar medium of galaxies, we observe it in a suite of low-ionization metals and atomic species (e.g., Si II, Mg II, CII, C I, Lyman series lines of H and D). This material coalesces into protostars and their surrounding protoplanetary disks where hot gas lines (e.g., C IV) serve as tracers of mass-accretion, and the dominant molecular species ($H_2$ and CO) are used to characterize the planet-forming environment. After these planets grow and emerge from their natal disks, their long-term stability can be probed in exospheric species such as H I, O I, and C II while the atmospheric markers for life on rocky planets in their respective habitable zones (HZs) are thought to be $O_2$, $O_3$, $CO_2$, and $CH_4$.

The common theme among all of these gas-phase diagnostics is that the strongest transitions (determined by their oscillator strengths) of the most abundant species (determined by their relative column densities) reside in the rest-frame ultraviolet bandpass, below the atmospheric cut-off at 3200 Å. Furthermore, with the exception of Ne VIII which must be redshifted above the Lyman limit at 912 Å and Mg II which is in the near-ultraviolet (NUV) at 2800 Å, *all* of these lines reside in the far-ultraviolet (FUV) bandpass from approximately 1000 – 1700 Å. For example, O VI ($\lambda\lambda$ 1032, 1038 Å) is a unique tracer of metal-rich gas at a temperature of approximately 300,000 K, HI Lyα (1216 Å) is the strongest line of the most abundant astrophysical species – ubiquitous in almost all astrophysical environments from the cosmic web to Earth's upper atmosphere. The Lyman and Werner ultraviolet band systems of $H_2$ have transition probabilities 15 – 18 *orders of magnitude* larger than the $H_2$ rovibrational transitions in the near- and mid-infrared, giving us direct access to the dominant molecular reservoir in most dense astrophysical environments without the need to resort to obscure isotopic variants of other molecules with uncertain conversion factors to the total gas mass. Even for the atmospheres of potentially habitable planets, the $O_3$ Hartley bands (2000 – 3000 Å) are by over an order of magnate the strongest ozone features in a terrestrial atmosphere. Therefore, the characterization of almost all astrophysical environments requires ultraviolet spectroscopy for a complete quantitative understanding. CHISL combines a high-resolution echelle mode with a moderate resolution imaging spectroscopy mode for a complete characterization of these environments.

## 2.1 Ultraviolet Spectroscopy from LUVOIR: The lifecycle of baryons, from the cosmic web to exoplanetary atmospheres

*Metals in the circumgalacitc medium around star-forming galaxies –*

Understanding the baryonic mass budget of the universe, how this gas flows into and out of galaxies, and how this gas flow regulates the star-formation history of the universe are key scientific questions that will be addressed with UV spectroscopy from the LUVOIR Surveyor. Models predict that a significant fraction of the "missing baryons" in the universe reside in a million-degree 'warm-hot IGM' that does not appear in intermediate-temperature IGM metal tracers such as O VI and C IV. The factor of ~100 increase in sensitivity (factor of ~25 increase in collecting area × factor of 4 lower instrumental background rates) of LUVOIR means that a huge number of background QSOs will be available for absorption line spectroscopy through this gas (down to AB magnitudes ~21 – 22 in the FUV, increasing the source density ~100 times relative to what is achievable with *HST;* Figure 1). This increased source density, the multi-object capability, and the high-sensitivity of the LUVOIR+CHISL will allow us to trace both the intermediate-temperature ($10^{5-6}$ K) and warm-hot ($10^{6-7}$ K) media simultaneously. The full complement of "coronal" gas lines ($\log_{10} T_{peak} > 10^{5.8}$; Ne VIII, Mg X, Na IX, Si XII) move into the LUVOIR FUV band between redshifts $z \approx 0.3 - 1$. LUVOIR + CHISL will provide effective areas and spectral resolutions several times larger than the most ambitious X-ray instruments, and in a portion of the QSO spectral energy distribution with more photons per second emitted than at X-ray wavelengths. In addition, very broad HI Lyα absorption lines can be used to probe million-degree material even in very low metallicity gas. Taken together, UV absorption line spectroscopy is the tool of choice for tracing the gas content of the IGM over the redshift range 0.3 – 3.0, through the major period of black hole and star-formation evolution.

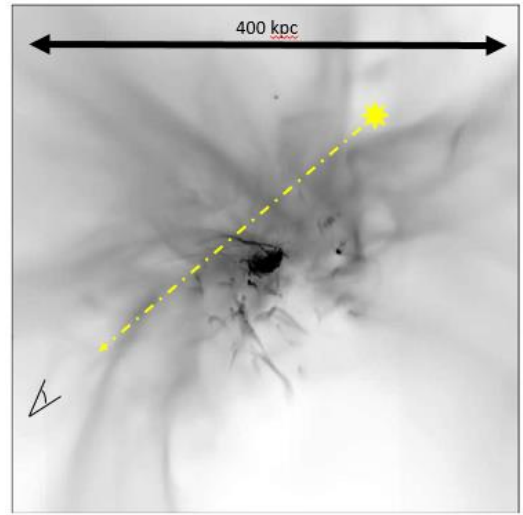

Figure 1 – CGM radial mass flow projection simulation from Joung et al. (2012; see also schematic in HDST report)[6]. The gray scale is mass flowing [g/cm/s] into a star-forming galaxy. The yellow sightline is a quasar absorption sightline though the CGM. LUVOIR will access enough background quasars to tomographically map the circumgalactic structure of low-to-intermediate z star-forming galaxies and may be able to map the emission from this gas directly.

Galaxy halos are the intersection points of the cosmic web, where dark matter concentrations focus the accumulation of baryons. This CGM is where we can study the inflow and outflow of gas into star-forming galaxies and understand the processes that regulate the star-formation history over approximately half of the age of the universe. The increased number of background QSOs noted above permit a full three-dimensional mapping (tomography) of the spatial distribution and velocity flows of material in the CGM of low-*z* star-forming galaxies. Furthermore, emission from this gas may be able to be mapped directly, creating a full three-dimensional picture of the CGM. While it is not yet clear if the densities and photon fluxes will be high enough for direct spectral imaging observations, this would truly revolutionize our understanding of the relationship between the baryon reservoirs around star-forming galaxies and the processes that transport that gas into and out of the sites of star and planet formation. Mapping the metal content, density, and temperature of the CGM enables us to quantify the amount of fuel flowing into galaxies to power ongoing star formation and how the infall rates compare to the amount of material being expelled from galaxies by galactic superwinds, driven by massive star clusters and supernovae[7-8]. The high velocity resolution of the CHISL echelle modes will enable the study of detailed velocity structure on specific sightlines through the CGM while the multi-object mode provides the three-dimensional mapping.

*Molecular Spectroscopy of Protoplanetary Disks: The Building Blocks of Exoplanets* –

Once this gas has been accreted by galaxies, it is eventually bound into dense molecular clouds and undergoes gravitational collapse to form protostellar systems. Roughly 1 Myr after onset of nuclear fusion in the protostar, the system emerges from its dense shroud of dust, and the raw materials for planet building can be measured with UV molecular absorption and emission line spectroscopy. The lifetime, spatial distribution, and composition of gas and dust in the inner $\sim$ 10 AU of young (age $\leq$ 30 Myr) circumstellar disks are important components for understanding the formation and evolution of extrasolar planetary systems. The composition and physical state (e.g., temperature, turbulent velocity, ionization state) of a cross-section of the circumstellar environment can be probed using high-resolution absorption line spectroscopy of high-inclination ($i > 60°$) disks. The bulk of the warm/cold $H_2$ gas is only observable at $\lambda < 1120$ Å (via the Lyman and Werner ($v'$ - 0) band systems). This work has only been possible on a small number of protoplanetary[9-11] and debris[12] disk systems to date. In addition to $H_2$, the strongest absorbing transitions of numerous species relevant to the atmospheres of both gas giant and rocky planets reside in the FUV, including CO, $CO_2$, $H_2O$, and $CH_4$ (Figure 2). Combining the high sensitivity (large collecting area + high-efficiency optical components, coatings, and detectors) of LUVOIR with the high-resolution and FUV bandpass of CHISL allows us to inventory all of these species simultaneously in a statistical sample of protoplanetary disks for the first time. This enables us to set the initial conditions for exoplanetary composition and atmospheres in unprecedented detail. In dense star-forming environments (e.g., Orion and more distant clusters), the multi-object capability of CHISL also enables simultaneous measurements of the molecular emission and absorption properties of disk surfaces regardless of whether or not the disk inclinations are suitable for high-resolution absorption line spectroscopy[13-14].

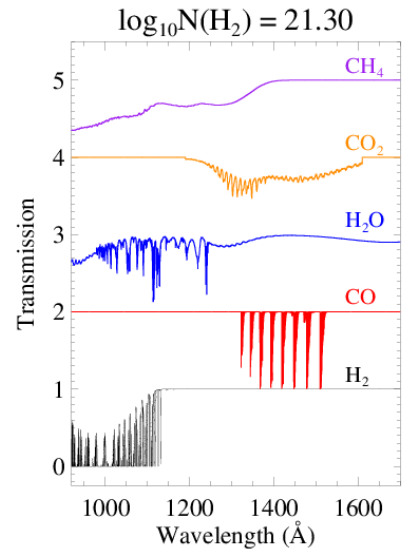

Figure 2 - Molecular absorption spectra of abundant species in the protoplanetary environment that are important constituents of terrestrial and gas giant atmospheres. High-sensitivity, high-resolution FUV spectroscopy can diagnose the composition, temperature, and stratification of all of these species in a single observations for moderately inclined disks.

*Long-term stability of Habitable Zone Atmospheres and the Formation of Biomarker Molecules –*

Once the planetary systems have formed and potentially habitable surface and atmospheric conditions have been established, the major challenge for exoplanet astronomers, and indeed the primary science focus for the LUVOIR mission, is the characterization of potential biomarker molecules in the atmospheres of Earth-like planets. The planetary effective surface temperature alone is insufficient to accurately interpret biosignature gases when they are observed by LUVOIR. The UV stellar spectrum drives and regulates the upper atmospheric heating and chemistry on Earth-like planets, is critical to the definition and interpretation of biosignature gases (e.g., Ref 15)[15], and may even produce false-positives in our search for biological activity[16-18] (Figure 3, right). High stellar irradiances in the FUV and EUV bands can lead to significant atmospheric mass-loss on timescales shorter than the time required for the development of a biosphere[19]. This is particularly true for rocky planets around low-mass stars where the HZ is at a fraction of the Earth-Sun separation and the relative fluxes in the X-ray, EUV, and FUV bands are significantly higher than for solar-type stars[20]. Therefore, our quest to observe and characterize biological signatures on rocky planets *must* consider the star-planet system as a whole, including the interaction between the stellar irradiance and the exoplanetary atmosphere.

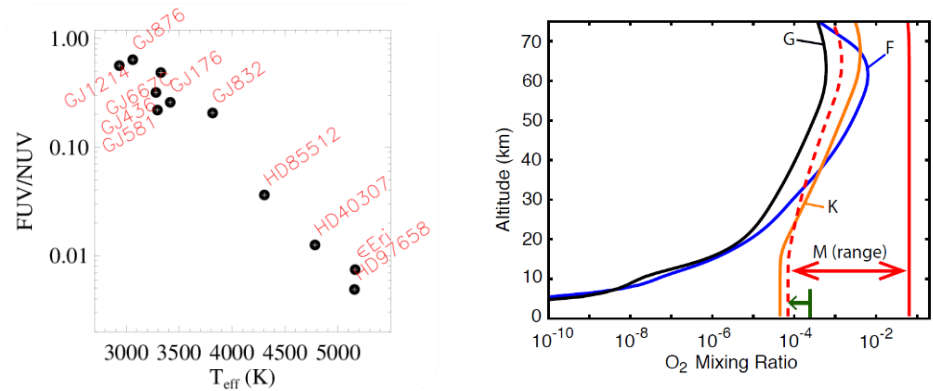

Figure 3 – (*left*) FUV/NUV flux ratio decreases dramatically as a function of the stellar effective temperature (from the MUSCLES Treasury Survey, Ref 20). (*right*) Large FUV/NUV flux ratios may lead to high concentrations of abiotic $O_2$ and $O_3$ on rocky planets orbiting M and K dwarfs. The plot at right shows the abiotic buildup of $O_2$ in the atmosphere of M stars, directly related to the FUV/NUV stellar flux ratio[40].

Spectral observations of $O_2$, $O_3$, $CH_4$, and $CO_2$, are expected to be important signatures of biological activity on planets with Earth-like atmospheres[21-23]. The chemistry of these molecules in the atmosphere of an Earth-like planet depends sensitively on the strength and shape of the host star's UV spectrum. $H_2O$, $CH_4$, and $CO_2$ are sensitive to far-UV radiation (FUV; 1000 – 1750 Å), in particular the bright HI Lyα line, while the atmospheric oxygen chemistry is driven by a combination of FUV and near-UV (NUV; 1750 – 3200 Å) radiation. While detailed observational and theoretical models are available for stars of solar-type and hotter, at present no theoretical models can predict the chromospheric, transition region, and coronal emission from an M or K star without a direct observation. Without knowledge of the stellar UV spectrum, the photoproduction rates for potential biomarker molecules cannot be accurately calculated and therefore the fraction of molecular species generated by disequilibrium processes such as life cannot be definitively assessed. In short, without direct observation of the UV spectra of low-mass exoplanet host stars, we will not be able to confidently assign the detection of potential biomarker molecules to biological processes. Therefore, UV spectroscopic characterization of the stars that LUVOIR will search for potentially habitable planets is critical to the interpretation of those results. Finally, given the much larger photoabsorption cross-sections of the relevant molecules at λ < 3000 Å, the FUV and NUV bandpasses may be promising spectral regimes for biomarker detection in their own right[24]. Access to the 2000 – 3200 Å NUV could be accommodated with the addition of an NUV mode, as described in Section 3.1.

Table 1 – NASA primary science goals addressed by LUVOIR

| NASA Key Science Goals | UV Spectroscopy | CHISL Data (HR = high-resolution, IS = imaging spectroscopy, MO = multi-object) |
|---|---|---|
| *Demographics of planetary systems* | | |
| *Characterizing other worlds* | X | Transit spectroscopy of terrestrial and Jovian planets, $O_2$ and $O_3$ edges (NUV required). [HR, IS] |
| *Our nearest neighbors and the search for life* | X | Characterization of host-stars: UV irradiance, flare frequency/amplitude distributions. Formation of biomarker molecules. [IS] |
| *The origins of stars and planets* | X | Molecules in protoplanetary disks (r < 10 AU, ages < 20 Myr), formation of planetary cores and atmospheres. [HR, IS, MO] |
| *The Milky Way and its neighbors* | X | Hot star populations and feedback (superwinds, supernovae), ISM of galaxies, circumgalactic halo emission and absorption [HR, IS, MO] |
| *The history of galaxies* | X | IGM absorption studies, CGM tomography and spectroscopic mapping, ionizing radiation escape fraction, He II Lyα forest [HR, IS, MO] |

## 3. THE COMBINED HIGH-RESOLUTION AND IMAGING SPECTROGRAPH FOR THE LUVOIR SURVEYOR: CHISL

The CHISL concept is designed to execute the LUVOIR science goals described in Sections 1 and 2. The LUVOIR science portfolio is broad, and as such several instrument modes will be required to fulfill the objectives of the mission. CHISL is highly multiplexed ultraviolet spectrograph, with both high-resolution and multi-object imaging spectroscopy modes. We envision as an analog to the successful *HST*-STIS instrument, with an order-of-magnitude higher efficiency. Our preliminary design also incorporates a microshutter array[41] (MSA; or other multi-slit capability) in the optical path, similar to NIRSPEC on *JWST*[42], to enable medium resolution multi-object spectroscopy ($R \approx 14000 - 35000$; N ≈ 700 apertures) over a 1′ × 2.4′ field-of-view.

CHISL makes use of pathfinder instrument designs vetted and tested on suborbital payloads (Section 4) with mid-to-high TRL component technology to enable order-of-magnitude gains in instrument throughput relative to comparable modes on *HST* (see, e.g., Table 2). Coupling the high instrumental throughput with a factor of ~25 gain in collecting area over HST ( $[12m / 2.4m]^2 = 25$), CHISL can reach to limiting fluxes of order 100 times fainter than currently possible. Alternatively, vastly larger samples of sources can be acquired in the same fixed amount of observing time. In Table 1, we briefly outline how CHISL supports each of NASA's Key Science Goals for LUVOIR noted in Section1. CHISL anticipates a moderately slow (> *f*/12) incident beam from an optical telescope assembly that has optical coatings compatible with high throughput broadband performance and wavefront control for visible-wavelength coronagraphy (e.g., Refs 25-27)[25-27]. CHISL will have a slit assembly for the moderate resolution imaging mode, as well as a secondary aperture for an echelle mode that will feed a collimating fore-optic.

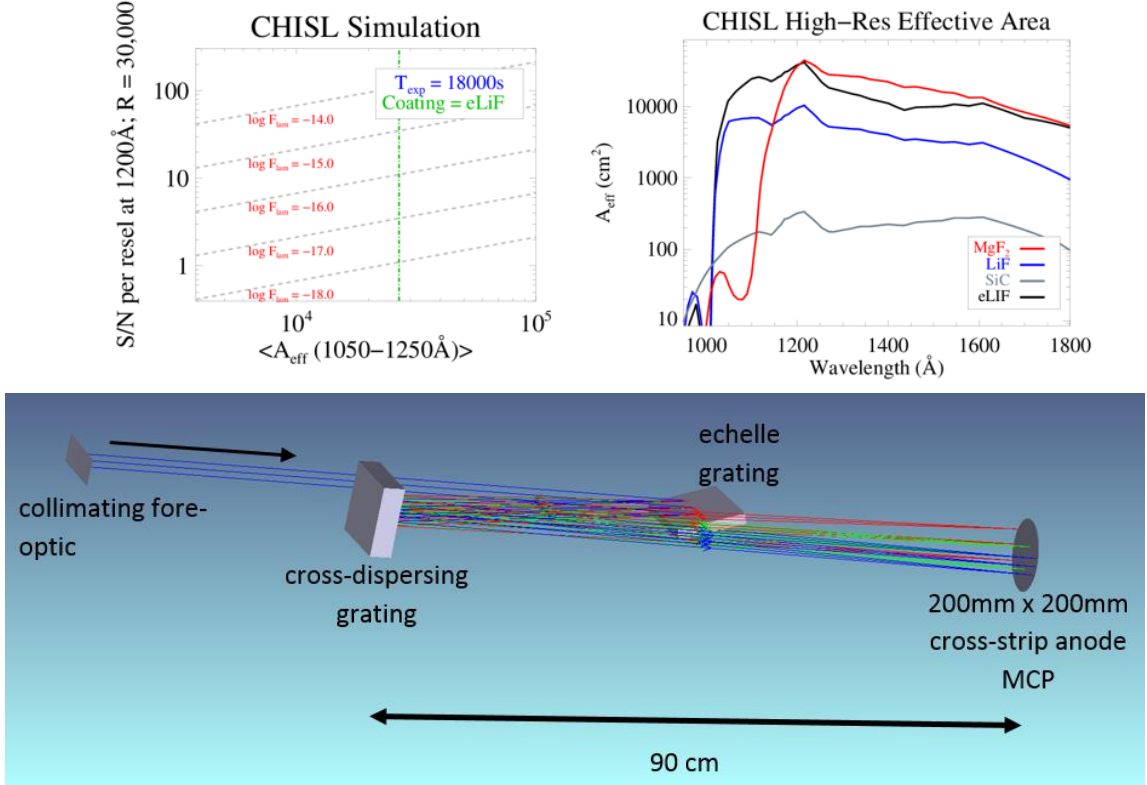

Figure 4 – (above) Raytrace of the CHISL high-resolution echelle mode ($R \approx 120{,}00$ from $1000 - 1700$ Å). At upper left and right, we present preliminary sensitivity and effective area curves (for a range of possible optical coatings), respectively. The curve at upper left presents S/N per resolution element for a 'typical' 18000s exposure (binned to $R = 30000$ for this calculation) as a function of source flux (in erg cm$^{-2}$ s$^{-1}$ Å$^{-1}$) and instrument effective area. Nominal effective area is shown as the green dashed line.

### 3.1 CHISL High-resolution and Imaging Spectroscopy Modes

*High-resolution* - The CHISL high-resolution spectroscopy mode operates at $R > 100{,}000$ over the $1000 - 1700$ Å bandpass. The 3 km s$^{-1}$ velocity resolution enables detailed studies of the mass flows and velocity fields in protoplanetary and circumgalactic environments while simultaneously providing separation of closely spaced rotational lines required for detailed quantitative molecular spectroscopy of species like CO. The CHISL detector arrays (nominally microchannel-plate arrays employing low-background borosilicate glass; Ref 28) have zero read noise and very low effective background flux, so high-resolution data can be readily binned to improve photon statistics for faint targets. We present a schematic instrument concept in Figure 4. Panchromatic light is fed by the LUVOIR optical telescope assembly, entering the spectrograph through an aperture offset from the multi-object array (Section 3.2) and collimated by a fore-optic. The collimated beam encounters a low-line density (~79 groove/mm) echelle grating operating at a high angle of incidence (~74°). Laboratory tests of similar gratings have found that these optics can achieve grove efficiency of order 70% in the FUV bandpass with modest (~ 10%) inter-order scatter. UV echelle gratings with improved groove efficiency and lower scatter should continue to be pursued as an enhancing technology for high-resolution UV spectroscopy.

The echelle directs the dispersed beam towards a toroidally shaped cross-dispersing grating ($R_1 \approx R_2 \approx 2500$ mm; ~350 groove/mm) which separates the orders and focuses the echellogram onto the detector focal plane. This design supports high-resolution across the 1000 – 1700 Å bandpass (the short wavelength cutoff is determined by the projected performance of the Al+eLiF optical coatings) and is relatively tolerant to misalignments and minor changes to the optical prescription of the gratings. The final spectrum consists of ~100 orders separated by 100 – 300 µm per order with a free spectral range of 4 – 7 Å per order being imaged onto the detector. The resolving power is detector limited, and assuming 25 µm spectral resolution elements achieves resolving power $R = 100{,}000 - 150{,}000$ across the 1000 – 1700 Å bandpass.

*Imaging Spectroscopy* - The CHISL long-slit or multi-object imaging spectroscopy mode consists of both low- and medium-resolution gratings on a grating wheel, with one slot open to allow for the possible future inclusion of an additional mode (e.g., an NUV mode feeding additional dispersive elements and an NUV-optimized detector, e.g., Nikzad et al. 2012). The medium resolution mode will be the primary mode, delivering $R > 24000$ over bandpasses of ~ 400 Å, while the low resolution mode provides $R \geq 14000$ in the range 1000 – 2000 Å. Both modes maintain $\Delta\theta \leq 200$ mas imaging in the cross-dispersion direction near the center of the FOV, degrading to ~ 1 arcsecond at the extrema. The gratings are blazed to 1200 Å and operate in first order near an angle of incidence of 2º with an approximately 2250 mm radius of curvature. They are both aberration corrected holographic gratings, with effective groove densities of 1200 and 1230 gr mm$^{-1}$ for the medium and low resolution modes, respectively. Image control is maintained with a relatively slowly converging beam that is extended by the inclusion of a toroidal fold mirror, making the total spectrograph focal length of 3.4 meters. The CHISL imaging field of view is 60″ x 144″. A schematic layout of the CHISL imaging spectroscopy channel, as well as preliminary spectral and angular resolution performance curves are shown in Figure 5.

**3.2 CHISL Optical Coatings and Detectors**

The working plan for the CHISL optics is to utilize "enhanced LiF" coatings, a high-temperature Al+LiF deposition technique developed as part of a NASA Strategic Astrophysics Technology grant at GSFC[25,27]. These coatings have demonstrated > 85% normal-incidence reflectivity over 1030 – 1300 Å, with sustained performance extending to NIR wavelengths. Testing and flight-qualification of these coatings is an ongoing area of research[27,29]. Advancements in ALD coating technology (e.g., Ref 30) will also be considered. The use of advanced coatings permits the inclusion of the fold mirror without a crippling loss of effective area for the CHISL imaging spectroscopy modes.

The CHISL spectrograms are recorded at the focal plane by two large format microchannel plate (MCP) detectors employing a CsI photocathodes. These detectors, each 200mm × 200mm MCPs, likely employing cross-strip anodes[31], record both the echellogram from the high-resolution mode (on one of the detectors) and the imaging spectral data from the imaging spectroscopy modes (on both detectors). The use of two primary science detectors increases the bandpass of the imaging spectroscopy mode and provides redundancy in the event that LUVOIR is not robotically serviceable. The large format, higher open area afforded by the new ALD borosilicate MCPs (10 – 15% larger open area than conventional MCP glass, with factors of ~3 – 4 lower particle backgrounds; Ref 28), and the optimized pore pitch (up to another 10% efficiency gain) have been enabled by an investment in the COR technology development. The use of an MCP device enables time-tagging capability for the investigation of transient and temporally variable astronomical phenomena (e.g. Ref 20). While the FUV performance of δ-doped CCDs[32] is currently less well established (see Refs 33 and 34 for early suborbital results), the capabilities and heritage of these detectors will be considered when the final CHISL instrument is defined. We will also consider separate detector systems for each "arm" of the spectrograph to increase redundancy.

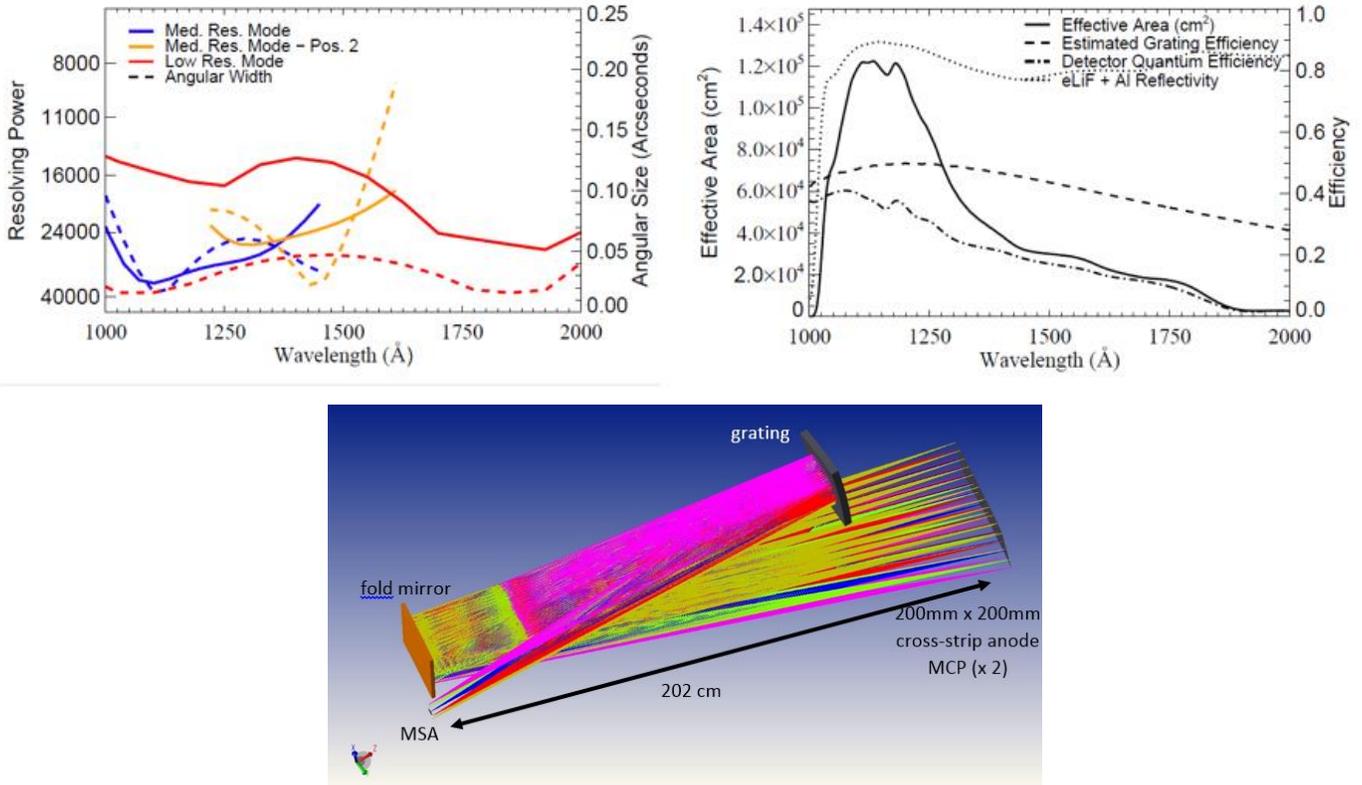

Figure 5. (above) Raytrace of the CHISL imaging spectroscopy mode. At upper left and right, we present preliminary performance and effective area curves, respectively. The curve at upper left shows the spectral resolving power (solid curves, left axis) for the low (red) and medium resolution (blue and orange) modes as well as the imaging performance for those modes (dashed lines, right axis). Nominal effective area, and assumed component efficiencies are shown at top right.

### 3.3 CHISL Simulated Data Products

We present a sketch of simulated LUVOIR + CHISL data products in Figure 6. The left hand panel in Figure 6 shows the simulated absorption line spectrum of a ~1 Myr protoplanetary system at a distance comparable to the Orion Nebula star-forming region. The simulation employs an H2 column density and temperature similar to the RW Aur protoplanetary disk ($\log_{10} N(H_2) = 20.2$, $T(H_2) \sim 400$ K)[43], scaled to approximately three times the distance of the Taurus-Auriga region and additional dust reddening. The signal-to-noise was estimated using the sensitivity curves presented in Figure 4. Water absorption was added based on the *Spitzer* and *Herschel* observations of protoplanetary disks showing that the CO/H$_2$O abundance ratios are ~1 in T Tauri star disks[44] (we assumed a fractional water abundance, $f(H_2O) = 5 \times 10^{-5}$). The spectra are shown here at $R = 30,000$ to match the calculations shown in Figure 4, but could be obtained at higher spectral resolution for higher-flux targets.

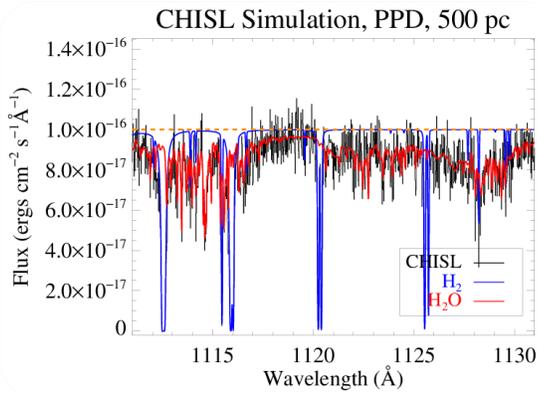

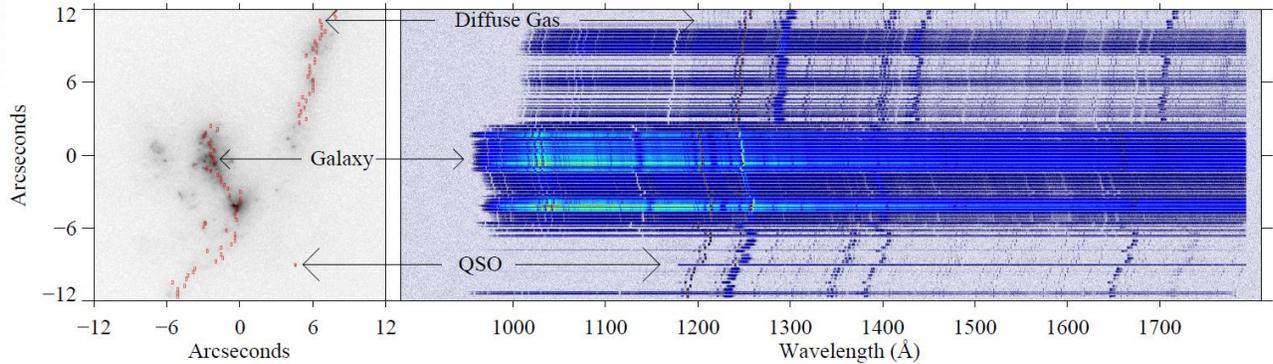

Figure 6 – (*left*) Simulated CHISL spectrum along a sightline through a high-inclination protoplanetary disk at the approximate distance of the Orion Nebula star-forming region. $H_2$ and $H_2O$ column density and temperatures can readily be measured, quantifying the initial abundances and physical state of the protoplanetary nebula. (*below*) CHISL multi-object spectroscopy of the disk + halo environment around a low-z star-forming galaxy. Quasar sightlines through the galaxy halo, direct CGM emission, and spectra of the galactic star-forming regions are all simulated from archival *HST*-COS and *FUSE* observations.

The right hand panels in Figure 6 show a "cartoon" of a multi-object spectrum of a low-redshift starburst galaxy. The multi-object capability enables multiple lines of sight to be probed simultaneously, including quasar sightlines through the halo (absorption line spectra), direct emission from the CGM (emission line spectra), and high S/N spectra of star-forming regions within the galaxy (a combination of emission and absorption lines diagnosing the stellar population and the state of the interstellar medium). Quasar and galactic sightlines are simulated using archival HST-COS spectra of these objects.

### 4. CHISL PATHFINDERS: ROCKET-BORNE INSTRUMENTS FOR TECHNOLOGY DEVELOPMENT AND FLIGHT-TESTING OF CRITICAL PATH COMPONENTS

The 2010 Decadal Survey recognized that many of the hardware components enabling a high-throughput multiplexed instrument for NASA's next UVOIR flagship would require significant technical development and called for an augmentation of NASA's budget for research and development of critical path technologies. NASA's Cosmic Origins Program office carries out a regular community assessment of technology needs for future missions and the COPAG SiG2 interest group on future flagships (see Section 1) identified technology drivers as well[5]. The University of Colorado's ultraviolet rocket program has a long history of carrying out the laboratory calibration and space-flight testing of hardware for NASA missions. Spectrograph and detector concepts developed at the University of Colorado have matured into large scale and highly successful NASA missions, including flight verification of MAMAs (in use on *HST*-STIS, -ACS, and the -COS NUV mode), virtual source holographically corrected UV gratings, and MCP delay-line detectors (enabling technology for both *FUSE* and *HST*-COS). This science and technology maturation program is carried out in the framework of a university program where undergraduate, graduate, and postdoctoral training is paramount. Much of the enabling hardware for the CHISL concept is currently undergoing laboratory and flight characterization aboard suborbital payloads at Colorado. We briefly describe the pathfinder instruments for the CHISL high-resolution and imaging spectroscopy channels: CHESS, the *Colorado High-resolution Echelle Stellar Spectrograph* and SISTINE, *Suborbital Imaging Spectrograph for Transition region Irradiance from Nearby Exoplanet host stars*.

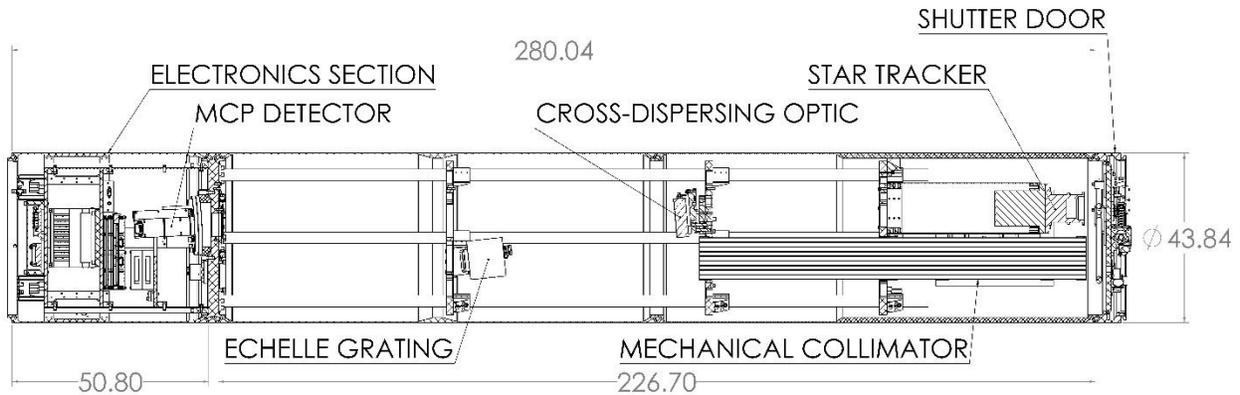

Figure 7. Schematic of the spectrograph section of the CHESS rocket payload. Labeled are relevant disk structures and optical components. All dimension in centimeters.

These payloads are designed to carry out unique astrophysical investigations and to respond to NASA's call for technology development and TRL maturation for future astrophysics missions, specifically the LUVOIR Surveyor. The COPAG SiG2 flagship study identified 1) high-efficiency optical coatings and 2) large format, high QE, photon-counting detectors as the top two technology needs to support a broad range of astrophysics with the LUVOIR Surveyor. NASA's 2014 COR Technology Report lists "High-Reflectivity Optical Coatings for UV/Vis/NIR" as the most important goal for Cosmic Origins (Item #1 on Priority 1 list). SISTINE will provide the first test of advanced Al+LiF coatings[25] on shaped optics and the first flight test of these coatings. Items #2 and #3 on the COR Priority 1 list both deal with large format and photon-counting UV detectors. CHESS is currently employing the new cross-strip MCPs developed as part of NASA's Strategic Astrophysics Technology (SAT) program, and will incorporate a next generation UV-sensitive δ-doped CCD detector on a future flight (also developed as part of the SAT program). The SISTINE MCP detector system will have a total active area of 220 mm × 40 mm; this large detector enables a factor of ten increase in the spectral bandpass and imaging field per exposure relative to the current state-of-the-art UV imaging spectrograph (*HST*-STIS, Table 2).

In this section, we present brief technical overviews of the CHESS and SISTINE payloads, laboratory characterization, and the flight-test schedule for the payloads. A more detailed discussion of these aspects can be found in Refs 35 – 37 and 27. An overview of the University of Colorado UV sounding rocket program and the suite of science payloads currently under development can be found in Ref 29.
*CHESS – The Colorado High-resolution Echelle Stellar Spectrograph*
CHESS is an objective echelle spectrograph operating at *f*/12.4 and resolving power of R ≈ 120,000 over a bandpass of 1000 – 1600 Å. CHESS is comprised of a mechanical collimator, mechanically ruled echelle grating, holographically ruled cross-disperser, and an FUV detector system. The spectrograph, detector, and supporting electronics are packaged into a standard 17.26" Wallops Flight Facility (WFF) provided rocket skin (Figure 7). The CHESS instrument has been described in previous technical conference proceedings[35,37]; we present a brief description here. The CHESS optical path can be described as follows (and see Figure 7):

- A mechanical collimator, consisting of an array of 10.74mm$^2$ × 1000mm anodized aluminum square tubes with a total collecting area of 40cm$^2$ and 0.9° field of view, prevents off-axis light from entering the spectrograph.
- An echelle grating (100 mm × 100 mm × 0.7 mm; Al+LiF-coated), with a groove density of 74 grooves mm$^{-1}$ and angle of incidence (AOI) of 63°, intercepts and disperses the stellar light. CHESS operates in orders m = 266 – 166. An initial design employed a 69 groove mm$^{-1}$ ruling density, however this optic proved impossible to fabricate at high efficiency. A follow-on electron-beam etched grating from JPL's MicroDevices Lab also could not be fabricated, and we are currently using a mechanically-ruled 53 grooves mm$^{-1}$ grating developed by Bach Research. At present, we are testing a 74 groove mm$^{-1}$ grating from Richardson for future flights of CHESS.
- A holographically ruled grating (developed in collaboration with J-Y Horiba and Al+LiF coated) is the cross dispersing optic (Figure 8). This grating is ion-etched into a laminar profile, using a deformable mirror in the recording solution, on a toroidal substrate for maximum first-order efficiency.
- A circular, 40 mm diameter cross-strip anode MCP detector[31] records the echellogram. This photon-counting device is capable of global count rates around 10$^6$ Hz.

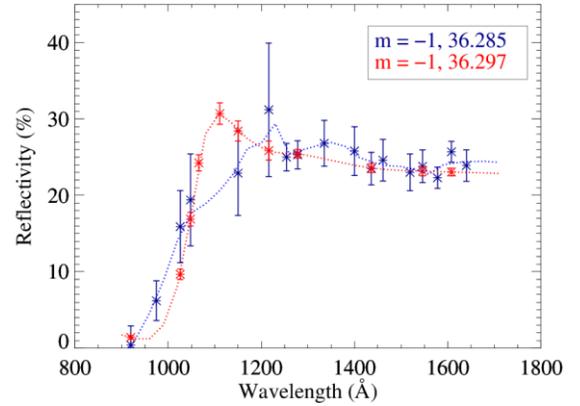

Figure 8. The CHESS cross-disperser reflectivity performance (note this is the total component reflectivity $G_{ech}(\lambda) = E_{ech}(\lambda) \times R_{ech}(\lambda)$). The Al+LiF coating shows no signs of degradation following 36.285 UG integration, launch, travel, and approximately nine months of dry storage.

The CHESS cross-strip MCP detector was built and optimized to meet the CHESS spectral resolution specifications at Sensor Sciences[39] (Figures 9 and 10). The detector has a circular format and a diameter of 40 mm. The microchannel plates are lead silicate glass, containing an array of 10-micron diameter channels. They are coated with an opaque cesium iodide (CsI) photocathode, which provides QE (35-15%) across the CHESS FUV bandpass (1000 – 1600 Å). There are two MCPs arranged in a "chevron" configuration. The detector achieves spatial resolution of 25 - 30 μm over an 8k pixels x 8k pixels format. While spatial resolutions over these size detectors have been achieved before with cross delay line anodes, the new cross strip format is capable of much higher global count rates, providing the first flight demonstration of a higher dynamic range MCP that will be essential for missions like LUVOIR that will have high photon arrival rates.

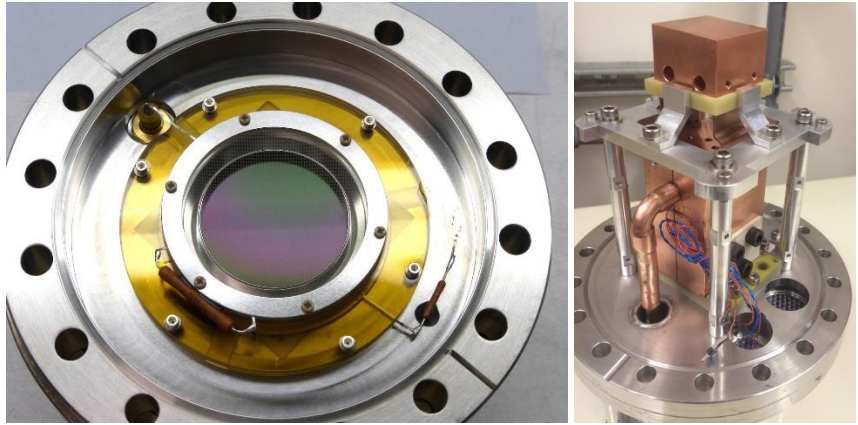

Figure 9. New hardware for the CHESS-2 and CHESS-3 payloads: *(left)* the 40mm cross-strip anode MCP detector following the deposition of a new photocathode at Sensor Sciences. CHESS-2 launched 21 February 2016 from White Sands Missile Range. *(right)* delta-doped CCD focal plane array electromechanical enclosure. The chip was developed at JPL and the focal plane and support electronics were built up at Arizona State University. The copper structure serves as the cold bath for the detector during flight. The CHESS flight employing the delta doped CCD is tentatively scheduled for a flight in mid-2017.

Following the launch of the CHESS-1, and in preparation for CHESS-2, we have had the 40mm cross-strip MCP refurbished at Sensor Sciences. This included the deposition of a new CsI photocathode, QE characterization, new digital output cables, and control systems functionality tests carried out in coordination with our team. A photo of the CHESS-2 detector, following refurbishment, is shown in Figure 9 (left). For a future flight of the CHESS payload (nominally scheduled for mid-2017), we will incorporate an FPA under development at ASU using a δ-doped CCD developed and provided by JPL (Figure 9. right). This device is a 3.5k pixels × 3.5k pixels, 10.5 μm pitch array providing ~30% DQE across the CHESS band. A first generation δ-doped device has been flown[33,34], however these CCD's modest format (330 × 1130 pixels with 24 μm pitch) will be improved upon significantly: the new CHESS CCD array expands the number of array elements by more than 30 times while reducing the pixel size by a factor of ≈ 2.3.

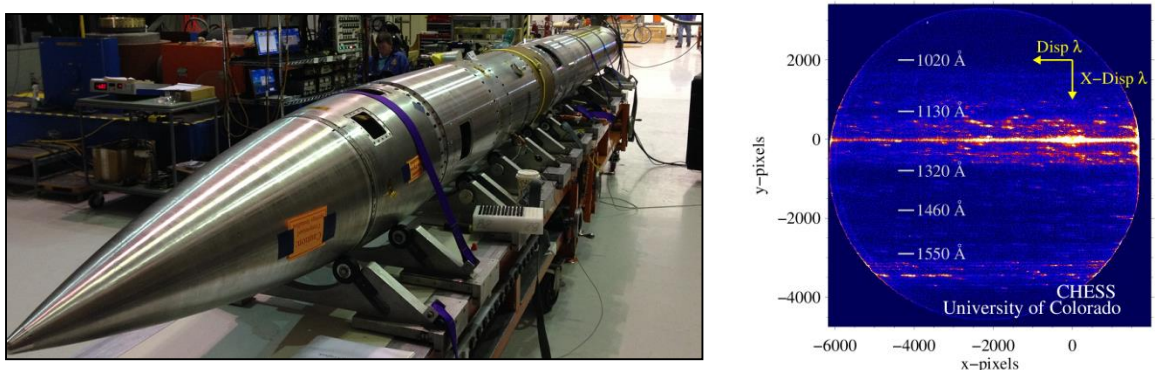

Figure 10 – (left) Built-up CHESS-1 payload during integration and testing of NASA/CU 36.285 UG (launch date: 24 May 2014). (right) Calibration data from the CHESS instrument, demonstrating R ≈ 70,000; the orders are separated top-to-bottom, each order has 5 – 7 Å of free spectral range.

The addition of the CCD FPA to the CHESS payload was accounted for in the early design of the instrument, so no major mechanical rebuild of the payload is required. This array is being tested at ASU at the time of writing (Figure 9, right) and is scheduled to be delivered to Colorado by mid-2016. The device allows enhanced sensitivity to $\lambda > 1400$ Å compared to CsI-coated MCPs and facilitates high-S/N UV observations that have proven to be difficult without fixed pixels in the detector system[39]. The CCD FPA will provide the first astronomical flight demonstration of a next-generation Si-based detector at far-UV wavelengths, allowing the performance of these devices to be assessed for inclusion in LUVOIR, while accommodating the science goals of the CHESS instrument.

*SISTINE – Suborbital Imaging Spectrograph for Transition region Irradiance from Nearby Exoplanet host stars*

At present, we are in the design phase of the SISTINE payload. SISTINE is an *f/*33 imaging spectrograph comprised of an *f/*16 classical Cassegrain telescope and a magnifying spectrograph. The system is designed for R ≈ 10,000 spectroscopy across the 1000 – 1600 Å bandpass with imaging performance between 0.5 – 2.0″ depending on the field position and wavelength (optimized for R ≈ 11,900 and Δθ ≈ 0.5″ at Lyα; Figure 11). SISTINE employs the new advanced Al+LiF ("eLiF", see above) coatings developed at GSFC[25] and is projected to deliver a peak effective area of $A_{eff}$ ≈ 170 cm$^2$ at Lyα, enabling high-sensitivity, moderate-resolution, astronomical imaging spectroscopy across this bandpass for the first time. SISTINE's *f/*16 telescope consists of a 0.5 m diameter parabolic primary mirror and a 86 mm convex hyperbolic secondary (Figure 11). The detector-area limited field-of-view is 6′. The primary mirror is *f/*2.5, and will be fabricated into a single-arch fused-silica substrate by Nu-Tek Corporation.

The SISTINE spectrograph is fed through a long-slit aperture (projected angular dimensions 5″ × 6′) at the telescope focus. The slit mount will be angled at 45º about the slit axis, polished, and optimized for visible reflectivity; the reflected image of the telescope field will be folded into a visible-light aspect camera to enable real-time maneuvering feedback during the flight. The light passing through the slit is dispersed by an aberration-corrected holographically-ruled spherical grating 745 mm from the slit toward the forward end of the rocket. The spherical figure will have a radius of curvature of 998 mm and is blazed to a peak efficiency at 1200 Å, with an effective ruling density of 1278 grooves mm$^{-1}$. The grating substrate is fused silica and will be fabricated by Precision Asphere, while the holographic ruling will be carried out by J-Y Horiba.

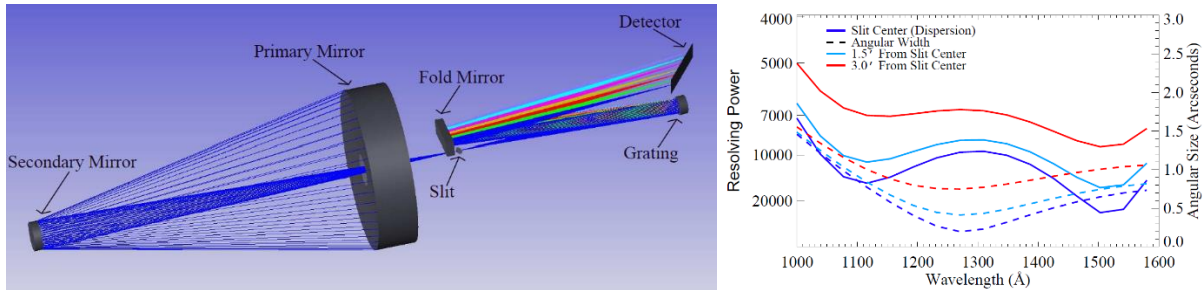

Figure 11. At left, we show a raytrace of the SISTINE payload. The telescope is a 0.5-m primary Cassegrain design. At right, we show performance projections for SISTINE. The solid lines show the predicted spectral resolving power (left axis) and the dotted lines show the imaging performance as a function of wavelengths and field position (in arcsec, right axis).

The use of eLiF coatings permits the inclusion of a fold mirror without compromising instrument performance. This allows the grating-detector distance to be larger than the slit-grating distance while fitting within the rocket envelope. This creates additional magnification in the spectrograph, enabling sufficient imaging resolution to reliably separate Lyα emission from the astronomical target from that introduced by airglow (and to angularly resolve nearby binary star systems). The long focal length also allows us to control aberrations and provide moderate resolution spectroscopy and imaging across large spectral (600 Å) and spatial ranges (6′) in a single observation (see Table 2 for a comparison of the instrument capabilities of SISTINE and *HST*-STIS' G140M medium resolution imaging spectroscopy mode).

Table 2 - Comparison between the CHISL imaging spectroscopy mode, SISTINE, and *HST*-STIS G140M

| *Instrument Parameter* | SISTINE | STIS G140M | CHISL (Imaging Modes) |
|---|---|---|---|
| Spectral Resolving Power | 10,000 | 10,000 | 16,000 – 40,000 |
| Total Spectral Bandpass | 1000 – 1600 Å | 1140 – 1740 Å | 1000 – 2000 Å |
| Spectral Bandpass per Exposure | 600 Å | 600 Å | 450 – 1000 Å |
| Number of Exposures to Cover Spectral Bandpass | 1 | 12 | 1 (Low Res) 3 (Med Res) |
| Imaging Field-of-View | 360″ | 28″ | 60″ x 144″ |
| Spectrograph Throughput | 14.1% | 1.2% | 11.7% |

The dispersed and imaged beam is recorded at the focal plane by a large format microchannel plane detector employing a CsI photocathode and a cross delay line (XDL) anode readout. The XDL MCP will consist of two 110 mm × 40 mm segments with a common anode, and will be purchased from Sensor Sciences, building on their extensive heritage of UV detectors for NASA's space missions. In a CHISL design, these detectors could be replaced by individual cross-strip anodes, possibly sharing the detector system with the high-resolution channel. For SISTINE, the detector will be mounted at a 30° angle relative to the incident beam in order to achieve the best focus and resolution while maintaining a flat focal plane, which minimizes costs. The XDL anode provides a photon-counting event list to enable the time-resolved spectroscopy. The large format, higher open area afforded by the new ALD borosilicate MCPs[28], and the optimized pore pitch support the COR technology priority for the development of higher-QE, large format, photon-counting UV detectors. SISTINE will provide the first flight test for these detector technologies in an end-to-end astronomical instrument.

## 5. SUMMARY

We have presented the first description of a multiplexed UV spectrograph for NASA's LUVOIR Surveyor mission concept, the *Combined High-resolution and Imaging Spectrograph for the LUVOIR Surveyor*. The two primary spectroscopic channels of CHISL were presented: a high-resolution echelle mode for $R \approx 120{,}000$ point source spectroscopy and an imaging spectroscopy mode operating in either a traditional long-slit or a multi-object spectral mode ($R = 14{,}000 - 35{,}000$ depending on the mode, FOV $\approx 1' \times 2.4'$). We have laid out three example science cases for extragalactic astrophysics and exoplanetary formation/characterization with the CHISL instrument. Initial instrument design considerations and data simulations were also presented. Both channels of the spectrograph rely on technology advancements that are realistic and under development at the time of writing: advanced UV coatings, high-QE large format photon-counting UV detectors, and low-scatter diffraction gratings operating at UV wavelengths. CHISL builds on the University of Colorado's ongoing laboratory and flight-testing technology programs, including a high-resolution spectrometer that has flown successfully on two NASA missions and an imaging spectrograph in development that is scheduled to fly twice before the 2020 Astronomy and Astrophysics Decadal Survey meets. This will enable us to incorporate the technology maturation from these suborbital payloads into the technical considerations made by NASA's STDTs for LUVOIR (and possibly HabEx).

*Acknowledgements* - This work was supported by NASA grant NNX13AF55G to the University of Colorado at Boulder. The authors thank Drs. Manuel Quijada, Paul Scowen, Shouleh Nikzad, John Vallerga, and Oswald Siegmund for their technical collaboration on the UV sounding rocket payloads described in this work. KF and BF also appreciate helpful discussions with Jason Tumlinson and Stephan McCandliss. The authors also acknowledge helpful suggestions made by an anonymous referee that have improved the quality of the final paper.

## REFERENCES

[1] 2010 Decadal Survey, "New Worlds, New Horizon's in Astronomy and Astrophysics", National Academies Press, 2010
[2] Postman, Marc, "Advanced Technology Large-Aperture Space Telescope (ATLAST): A Technology Roadmap for the Next Decade", Astro2010, arXiv:0904.0941, 2009
[3] NASA, "2013 Astrophysics Roadmap: Enduring Quests, Daring Visions: NASA Astrophysics in the next Three Decades", 2013
[4] Seager, S., et al., "From Cosmic Birth to Living Earth: A visionary space telescope for UV-Optical-NearIR Astronomy", AURA, http://www.hdstvision.org/report/, 2015
[5] Sembach, K., et al. "Cosmic Origins Program Analysis Group (COPAG) Report to Paul Hertz Regarding Large Mission Concepts to Study for the 2020 Decadal Survey", NASA COPAG, http://cor.gsfc.nasa.gov/copag/rfi/, 2015
[6] Joung, M. Ryan; Putman, Mary E.; Bryan, Greg L.; Fernández, Ximena; Peek, J. E. G., "Gas Accretion is Dominated by Warm Ionized Gas in Milky Way Mass Galaxies at z ~ 0", ApJ, 759, 137, 2012.
[7] Heckman, Timothy M.; Armus, Lee; Miley, George K. "On the nature and implications of starburst-driven galactic superwinds", ApJS, 74, 833, 1990
[8] Dawson, J. R, et al. "Supergiant Shells and Molecular Cloud Formation in the Large Magellanic Cloud", ApJ, 763, 56, 2013

[9] Roberge, A., et al. "FUSE and Hubble Space Telescope/STIS Observations of Hot and Cold Gas in the AB Aurigae System", ApJL, 551, 97, 2001
[10] France, Kevin, et al. "CO and $H_2$ Absorption in the AA Tauri Circumstellar Disk", ApJ, 743, 186, 2012c
[11] France, Kevin; Herczeg, Gregory J.; McJunkin, Matthew; Penton, Steven V. "CO/$H_2$ Abundance Ratio ≈ $10^{-4}$ in a Protoplanetary Disk", ApJ, 794, 160, 2014
[12] Roberge, A., et al. "High-Resolution Hubble Space Telescope STIS Spectra of C I and CO in the β Pictoris Circumstellar Disk", ApJ, 538, 904, 2000
[13] France, Kevin, et al. "The Far-ultraviolet "Continuum" in Protoplanetary Disk Systems. II. Carbon Monoxide Fourth Positive Emission and Absorption", ApJ, 734, 31, 2011
[14] France, Kevin, et al. "A Hubble Space Telescope Survey of $H_2$ Emission in the Circumstellar Environments of Young Stars", ApJ, 756, 171, 2012
[15] Seager, S.; Bains, W.; Hu, R., "Biosignature Gases in $H_2$-dominated Atmospheres on Rocky Exoplanets", APJ, 777, 95, 2013
[16] Hu, Renyu; Seager, Sara; Bains, William. "Photochemistry in Terrestrial Exoplanet Atmospheres. I. Photochemistry Model and Benchmark Cases", ApJ, 761, 166, 2012
[17] Tian, Feng, et al., 2014, "High stellar FUV/NUV ratio and oxygen contents in the atmospheres of potentially habitable planets", E&PSL, 385, 22
[18] Domagal-Goldman, Shawn D., et al, 2014, "Abiotic Ozone and Oxygen in Atmospheres Similar to Prebiotic Earth", ApJ, 792, 90
[19] Lammer, H., et al., "Determining the mass loss limit for close-in exoplanets: what can we learn from transit observations?", A&A, 506, 399, 2009
[20] France, Kevin, et al. "The MUSCLES Treasury Survey: Motivation and Overview", ApJ, *submitted*, 2016b
[21] Des Marais, David J., et al., "Remote Sensing of Planetary Properties and Biosignatures on Extrasolar Terrestrial Planets", AsBio, 2, 153, 2002
[22] Kaltenegger, Lisa, Traub, Wesley A., & Jucks, Kenneth W., "Spectral Evolution of an Earth-like Planet", ApJ, 658, 598, 2007
[23] Seager, S., Deming, D., & Valenti, J. A., "Transiting Exoplanets with JWST", Astrophysics in the Next Decade, Astrophysics and Space Science Proceedings, Springer, 123, 2009
[24] Bétrémieux, Y. & Kaltenegger, L., "Transmission Spectrum of Earth as a Transiting Exoplanet from the Ultraviolet to the Near-infrared", ApJ, 772, 31, 2013
[25] Quijada, M. et al. "Enhanced far-ultraviolet reflectance of $MgF_2$ and LiF over-coated Al mirrors", SPIE, v9144, 2014
[26] Balasubramanian, Kunjithapatham, et al., "Aluminum mirror coatings for UVOIR telescope optics including the far UV", SPIE, 9602, 01, 2015
[27] Fleming, Brian T., et al., "New UV instrumentation enabled by enhanced broadband reflectivity lithium fluoride coatings", 9601, 0R, 2015
[28] Siegmund, O. H. W, et al., "Advances in microchannel plates and photocathodes for ultraviolet photon counting detectors", SPIE, 8145, 0J, 2011
[29] France, Kevin, et al. "The SLICE, CHESS, and SISTINE Ultraviolet Spectrographs: Rocket-borne Instrumentation Supporting Future Astrophysics Missions", JAI, *in press*, 2016
[30] Moore, Christopher S.; Hennessy, John; Jewell, April D.; Nikzad, Shouleh; France, Kevin,"Recent developments and results of new ultraviolet reflective mirror coatings", SPIE, 9144, 4, 2014
[31] Vallerga, John, et al. "Cross strip anode readouts for large format, photon counting microchannel plate detectors: developing flight qualified prototypes of the detector and electronics", SPIE, 9144, 3, 2014
[32] Nikzad, Shouleh, et al., "Delta-doped electron-multiplied CCD with absolute quantum efficiency over 50% in the near to far ultraviolet range for single photon counting applications", ApOpt, 51, 365, 2012
[33] France, K.; Andersson, B.-G.; McCandliss, S. R.; Feldman, P. D., "Fluorescent Molecular Hydrogen Emission in IC 63: FUSE, Hopkins Ultraviolet Telescope, and Rocket Observations", ApJ, 628, 750, 2005
[34] Lupu, Roxana E.; McCandliss, Stephan R.; Fleming, Brian; France, Kevin; Feldman, Paul D.; Nikzad, Shouleh "Calibration and flight performance of the long-slit imaging dual order spectrograph", SPIE, 7011, 20
[35] France, Kevin; Beasley, Matthew; Kane, Robert; Nell, Nicholas; Burgh, Eric B.; Green, James C. "Development of the Colorado High-resolution Echelle Stellar Spectrograph (CHESS)", 8443, 05, 2012
[36] France, K. et al. "Flight performance and first results from the sub-orbital local interstellar cloud experiment (SLICE)", SPIE, 8859, 10, 2013b

[37] Hoadley, K., France, K. et al. "The assembly, calibration, and preliminary results from the     Colorado high-resolution Echelle stellar spectrograph (CHESS)", SPIE, v9144, 2014

[38] Siegmund, Oswald H. W.; Tremsin, Anton S.; Vallerga, John V. "Development of cross strip MCP detectors for UV and optical instruments" SPIE, 7435, 0L, 2009

[39] Ake, T. B. et al. "COS FUV Flat Fields and Signal-to-Noise Characteristics", HST Calibration Workshop, 2010

[40] Harman, C. E.; Schwieterman, E. W.; Schottelkotte, J. C.; Kasting, J. F. "Abiotic $O_2$ Levels on Planets around F, G, K, and M Stars: Possible False Positives for Life?", ApJ, 812, 137, 2015

[41] Li, M. J.; Brown, A. D.; Kutyrev, A. S.; Moseley, H. S.; Mikula, V. "JWST microshutter array system and beyond", SPIE, 7594, 75940, 2010

[42] Kutyrev, A. S.; Collins, N.; Chambers, J.; Moseley, S. H.; Rapchun, D. "Microshutter arrays: high contrast programmable field masks for JWST NIRSpec", SPIE, 7010, 70103, 2008

[43] France., K.; Herczeg, G., McJunkin, M., and Penton, S. "CO/$H_2$ Abundance Ratio Observed in a Protoplanetary Disk", ApJ, 794, 160, 2014

[44] Carr, J. & Najita, J. "Organic Molecules and Water in the Inner Regions of T Tauri Stars", ApJ, 733, 102, 2011